\documentclass{article}

\usepackage[english]{babel}
\usepackage{siunitx}
\usepackage[a4paper,top=2cm,bottom=2cm,left=3cm,right=3cm,marginparwidth=1.75cm]{geometry}

\usepackage{authblk}
\usepackage{amsmath}
\usepackage{graphicx}
\usepackage[colorlinks=true, allcolors=blue]{hyperref}
\usepackage{wrapfig}
\usepackage[capitalise]{cleveref}
\usepackage{makecell}
\usepackage{multicol}
\usepackage{wrapfig}
\usepackage[backend=bibtex,style=phys]{biblatex}
\usepackage{caption}
\usepackage{multirow}
\usepackage{booktabs}
\usepackage{customtitle}
\usepackage[T1]{fontenc}
\DeclareSIUnit\angstrom{\text {Å}}

\bibliography{bib.bib}

\title{Njord and Remora}
\subtitle{Spectrometers in Symbiosis}

\author[1]{E.~Fogh}
\author[2]{N.~L.~Amin}
\author[2]{G.~S.~Tucker}
\author[2]{M.~Aouane}
\author[1,3]{R.~Georgii}
\author[4]{J.~Voigt}
\author[2,5]{R.~Toft-Petersen}
\affil[1]{Technische Universit\"at M\"unchen}
\affil[2]{European Spallation Source ERIC}
\affil[3]{Heinz Maier-Leibnitz Zentrum}
\affil[4]{Forschungzentrum J\"ulich}
\affil[5]{Danmarks Tekniske Universitet}

\begin{document}
\settitlebefore{1.5em}
\settitleupper{0.5em}
\settitlelower{4.5em}
\settitleafter{1em}
\settitleleft{0.45\textwidth}
\settitleright{0.52\textwidth}
\maketitlecustom

\section{Executive summary}

Many of the most interesting scientific subjects are also the hardest to study with neutrons. Metal-organic frameworks, organic superconductors, quantum magnets, pressure-tuned materials, are systems where the relevant signals are weak, the samples are tiny, or the experiments need extreme sample environments such as pressure cells and high-field cryomagnets. Existing instruments often run into practical limits before the science is exhausted. For some questions the samples are simply too small; for others, the signal is buried in background or the required measurement time becomes prohibitive. This is both a scientific opportunity and a challenge for the European neutron scattering community.

\medskip\noindent
We present \textbf{Njord} and \textbf{Remora} as a paired instrument concept for the European Spallation Source (ESS).
The proposal focuses on two linked problems: important science cases are being limited by neutron flux and sample geometry, and the community also needs more beamtime. Njord addresses the first by pushing the available brightness into a tightly focused beam, while Remora uses the remaining spectral window to add a complementary spectrometer on the same beamport. A website at \href{https://njordremora.org/proposal/njord-and-remora-2026/}{njordremora.org} has been created for the instrument duo with further information and a showcase of community engagement.

\medskip\noindent
\textbf{Njord} will allow collective excitations to be measured on sub-mm$^3$ sample volumes, which would not be possible today due to insufficient count rates, even at the ESS. Njord aggressively focuses the intensity from the worlds most intense neutron source, by using the novel Nested-Mirror-Optics (NMO) technology, and measures neutrons scattered into a large solid angle with a large out-of-plane coverage. The design is optimised to exploit the full ESS pulse. As a result, Njord will deliver a small, well-defined beam (3x3 mm$^2$) with a flux exceeding $2.5 \times 10^{10}$ n/s/cm$^2$ (at 2 MW), while retaining ample space for sample environment equipment. Relaxing the resolution of the secondary spectrometer, the full scattering angle range will be covered continuously. Thus optimised for high flux, extreme-conditions studies, and out-of-plane coverage, Njord targets the experimental gap described above, enabling investigations of magnetic systems inaccessible to neutron spectroscopy today. Together with BIFROST, Njord represents a significant step towards maintaining ESS as a long-term global leader in high-flux spectrometer instrumentation.

\medskip\noindent
To address the capacity challenge, we propose \textbf{Remora}. This direct-geometry spectrometer will be positioned upstream of Njord on the same beamport and utilises the part of the neutron spectrum that cannot be accepted by Njord. It requires only a few standard components to shape and monochromate the incoming beam as well as a compact secondary spectrometer. The result is a direct-geometry spectrometer with a large dynamic range enabled through higher order reflections and a flux exceeding 10$^5$ n/s/cm$^2$.

\medskip\noindent
The combination of \textbf{Njord} and \textbf{Remora} delivers a two-pronged advancement to the ESS mission:
\begin{itemize}
    \item[] \textbf{Expanding scientific scope:} Enabling revolutionary experiments on challenging magnetic and organic systems, previously beyond the scope of neutron spectroscopy.
    \item[] \textbf{Increasing capacity and access:} Providing crucial beam time and user access, thus supporting a diverse, dynamic, and sustainable spectroscopy community.
\end{itemize}
We believe this symbiotic instrument duo will play a pivotal role in realising ESS’s vision as the leading neutron source in Europe, both in instrument capability and community impact.

\section{Scientific case}
Njord stands out on two main points: (i) extreme flux on a very small beam spot size, and (ii) large out-of-plane analyser coverage. These characteristics address a large and varied user community spanning from the more traditional hard condensed matter cases to planetary physics and organic matter. Njord brings the opportunity to study the following:
\begin{itemize}
    \item[-] Materials that inherently grow in the form of small crystals, like those obtained via high+pressure synthesis, flux growth or in organic networks. Low-energy collective dynamics in such materials are widely unexplored.
    \item[-] Studies under high uniaxial or hydrostatic pressures, in particular studies using diamond-anvil or strain cells become feasible.
    \item[-] Horizontal-magnetic-field studies on small samples with out-of-plane coverage.
    \item[-] Small-sample powder spectroscopy, where the out-of-plane coverage allows efficient integration over the Debye-Scherrer cone.
    \end{itemize}
In Table \ref{tab:fields} we list the various relevant research fields and below we highlight a few specific examples. We focus on the novel capabilities provided by Njord and only briefly point out the added capacity provided by Remora as a standard time-of-flight (ToF) direct-geometry spectrometer.

\begin{table}[h]
    \centering
    \caption{Research areas addressed by Njord listed in alphabetic order. Key examples are described in more detail below. }
    \label{tab:fields}
    \begin{tabular}{lp{10cm}}
	 \toprule
         Research field        & Examples \\
         \midrule
         Chemistry              & In-situ high-pressure synthesis, new materials \\
         Energy and environment & Barocalorics, batteries, catalysis, metal-organic frameworks \\
         Life sciences and soft matter         & Lamellar phases, liquid crystals, membranes, proteins \\
         Planetary physics      & Clathrates, hydrogen hydrates, ice systems \\
         Quantum phenomena & Altermagnets, heavy-fermion metals, metal-insulator transitions, quantum magnets,  superconductors, topological materials \\
	 \bottomrule
    \end{tabular}
\end{table}

\subsection{Metal-organic frameworks}
Materials science is applying the knowledge gained in fundamental research to design functional materials for the benefit of society. However, in simple ionic bonded materials, such efforts often come with unforeseen complexities inherent in realised materials. In this regard, metal–organic frameworks (MOFs) constitute a rapidly expanding class of crystalline hybrid materials in which metal nodes and organic linkers assemble into highly tunable lattices, allowing material engineering at the molecular level, including magnetic interactions and lattice dynamics \cite{furukawa2013,alhamami2014}. Their importance was underlined by the 2025 Nobel Prize in Chemistry.

\begin{wrapfigure}{r}{0.4\textwidth}
    \includegraphics[width=0.4\textwidth]{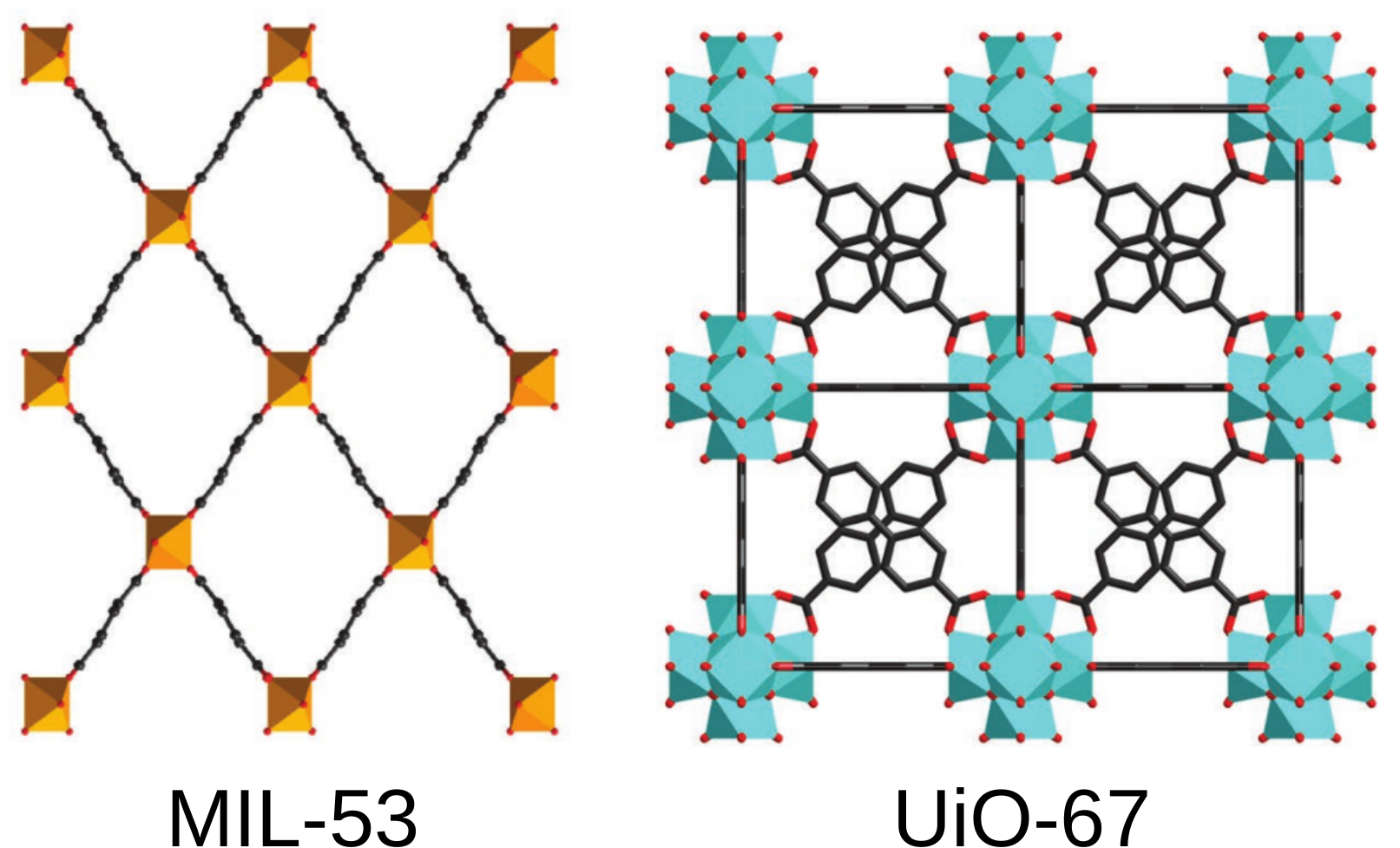}
    \caption{The structure of the MIL-53 and UiO-67 MOFs \cite{yuan2018}.}
    \label{fig:mofs}
\end{wrapfigure}

One example is known as the breathing MOF \cite{loiseau2004}, exemplified by what is known as the MIL-53 system (Fig. \ref{fig:mofs}). When absorbing gas or as a function of temperature or pressure it undergoes a reversible structural transition, where the unit cell volume changes by almost 40 \% and hence behaves almost like a lung, i.e., changeable yet stable. Vibrational spectroscopy and DFT shed light on the ${\bf Q} = 0$ modes \cite{kolokolov2010}, yet the collective modes driving the structural instability have not yet been resolved and here the full $\bf Q$-resolved dispersion is key. The tunability of the breathing MOFs enable applications as hydrogen storage materials, gas purifiers, water treatment and within controlled drug delivery. Other MOFs such as the UiO-67 (Fig. \ref{fig:mofs}) may be used for photocatalysis \cite{yuan2018}.

The chemical flexibility of MOFs further allows for incorporation of magnetic ions and thereby modular tailoring of quantum magnetic compounds for realising theoretical models. Traditional hard-condensed matter model compounds are rare to come by, and may be difficult to manipulate into the interesting parameter regime using external stimuli such as pressure or magnetic fields. Therefore, MOFs offer an attractive parallel path to studies of clean quantum magnetic systems.

A prominent example is $\kappa$-(BEDT-TTF)$_2X$, where (BEDT-TTF) is the organic molecule C$_{10}$H$_8$S$_8$, and $X$ is an inorganic unit involving Cu, F, I, Cl and/or Br \cite{commeau2017,riedl2022}. This family spans antiferromagnetic, superconducting and semiconducting states as a function of pressure \cite{dumm2009}, with a ground state highly sensitive to pressure and magnetic field \cite{kagawa2004,kagawa2005}. Two notable examples are the identification of a spin-liquid state in $\kappa$-(BEDT-TTF)$_2$Cu$_2$(CN)$_3$ \cite{isono2016,pustogow2022}, and the discovery of spin-polarised currents in $\kappa$-(BEDT-TTF)$_2$Cu[N(CN)$_2$]Cl \cite{naka2019}. The magnetic interaction network resembles that of what is known as the Shastry-Sutherland lattice \cite{shastry1981}, hosting extraordinarily rich quantum physics due to this geometric spin arrangement \cite{shastry1981,corboz2013,miyahara1999,miyahara2003,nomura2023,yang2022}. $\alpha$- and $\kappa$-(BEDT-TTF)$_2X$ systems were suggested as physical realisations for altermagnetism on the Shastry-Sutherland lattice, offering a path towards altermagnetism within a frustrated lattice. The possibility for a link between altermagnetism and superconductivity renders these systems of particular interest.

\medskip\noindent
Due to the complexity of the MOFs, large single crystals are inherently hard to come by \cite{delrieu1987,gandara2014}, rendering investigations of collective excitations extremely challenging.  Due to the porous nature of the MOFs, the lattice and vibrational modes have low energies. X-ray spectroscopy is therefore rarely the right tool, both due to the energy resolution and the low sensitivity to hydrogen. While vibrational neutron spectroscopy has established that low-energy linker motions dominate the dynamics of flexible metal–organic frameworks \cite{kolokolov2010,ryder2014,ryder2017}, identifying whether these modes are localised molecular vibrations or collective lattice phonons remains challenging. Since collective phonons control the elastic stability, breathing transitions, and thermodynamic properties of the framework, benchmarking phonon measurements against DFT calculations is crucial.

\subsection{Organic superconductors}

\begin{wrapfigure}{r}{0.5\textwidth}
    \vspace{-10mm}
    \includegraphics[width=0.5\textwidth]{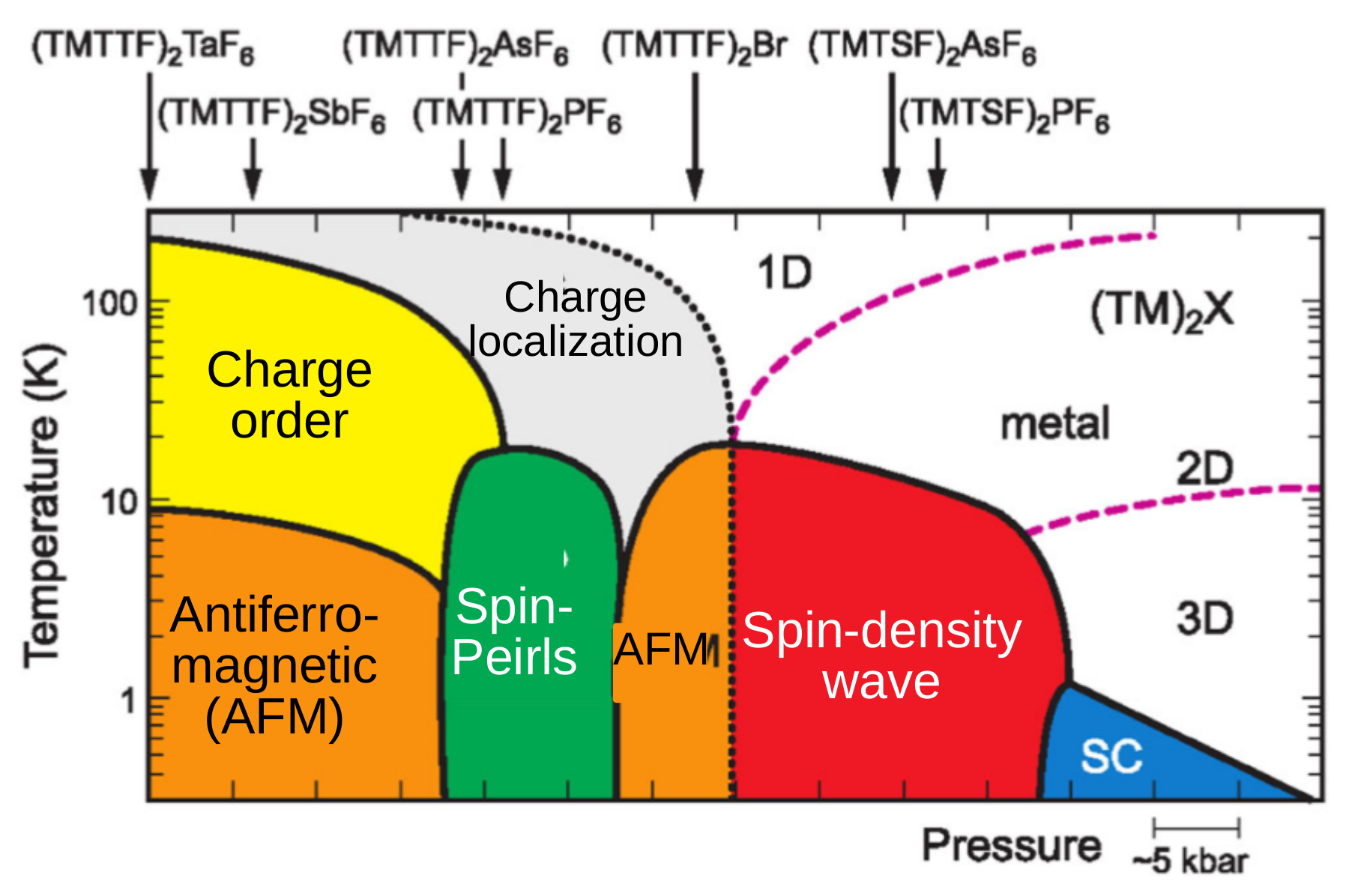}
    \caption{
    Pressure-temperature diagram of the organic 1D superconductors \cite{wang2023} Each material resides on this phase diagram and thus the pressure axis is not on an absolute scale.}
    \label{fig:salts}
\end{wrapfigure}

Superconductivity remains one of the most challenging and exciting areas of condensed matter physics. Traditional superconductors are well understood, where lattice vibrations (phonons) mediate electron pairing, as described in the Bardeen-Cooper-Schrieffer (BCS) theory \cite{BCStheorypaper}. Unconventional superconductivity arising from magnetic parent states have been a major focus of cold neutron spectroscopy for decades. The cuprates \cite{lee2006}, discovered in 1986 \cite{bednorzmuller1986}, are doped Mott insulators with an antiferromagnetic parent state, while the pnictides \cite{johnston2010}, discovered in 2008 \cite{kamihara2008}, are bad metals with itinerant magnetism. However, the first discovery of unconventional superconductivity in magnetic materials occurred earlier, with the 1980 identification of Bechgaard charge-transfer salts \cite{jerome1980,jerome2002}. Here, the magnetism resides in the molecular orbitals, rather than the atomic ones. The orbitals are delocalised, and facilitate electron itinerancy mainly of 1D nature along the stack, via Fermi surface nesting. The phase diagram of the organic-superconductor class is shown in Fig. \ref{fig:salts}, and is exceedingly complex. Depending on the inter-molecule stacking separation, the systems can be tuned to host charge order, antiferromagnetism, low dimensional metallic states, a Spin-Peierls instability and finally a spin-density wave preceding superconductivity.

Like the cuprates and pnictides, spin fluctuations are thought to be the microscopic mechanism giving rise to the superconducting state in organic superconductors. Here, however, superconductivity emerges from a 1D spin-density wave, and the fundamental energy scales are much smaller. The organic superconductors can be remarkably cleanly tuned, via chemical doping and pressure, allowing the full phase diagram to be studied. Although there is considerable agreement that the superconducting mechanism is magnetic in origin, there is yet to be decisive confirmation and an established theory of both superconductivity and the metal-dimensionality crossover in the organic superconductors. This is reminiscent of the situation in the cuprates, where a theory of the superconducting mechanism has yet to the confirmed. The organic superconductors, however, are simpler and more tunable than the cuprates, and offer a cleaner testing ground for pairing-mechanism theories. Therefore, understanding superconductivity in organic superconductors is the crucial next step toward an all-embracing theory of unconventional superconductivity.

\medskip\noindent
For neutron studies of organic superconductors, the reduced form factor of delocalised moments, and the frequent requirement for hydrostatic pressure to tune the system pose additional challenges. Consequently, the collective spin fluctuation spectrum of these systems remains largely unknown. Exactly these fluctuations are thought to drive superconductivity as well as provide evidence as to their magnetic state. Njord is designed precisely for such cases and will enable the first momentum- and energy-resolved studies as well as studies under pressure of spin excitations in organic superconductors.

\subsection{Powders under pressure}

One particular advantage of Njord is the very large spatial angle coverage of the analyser system, which renders the spectrometer ideally suited to investigate low-energy excitations in small amounts of powder samples. In contrast to in-plane-limited indirect-geometry spectrometers, large volumes of $S({\bf Q}, \omega)$ around the Debye-Scherrer cone can be measured simultaneously with Njord.
The focusing optics of Njord minimise illumination of sample environment components while providing extreme flux on the sample; this gives extraordinary new opportunities for pressure studies of materials. Examples are given below.

\begin{wrapfigure}{r}{0.55\textwidth}
    \vspace{-5mm}
    \includegraphics[width=0.55\textwidth]{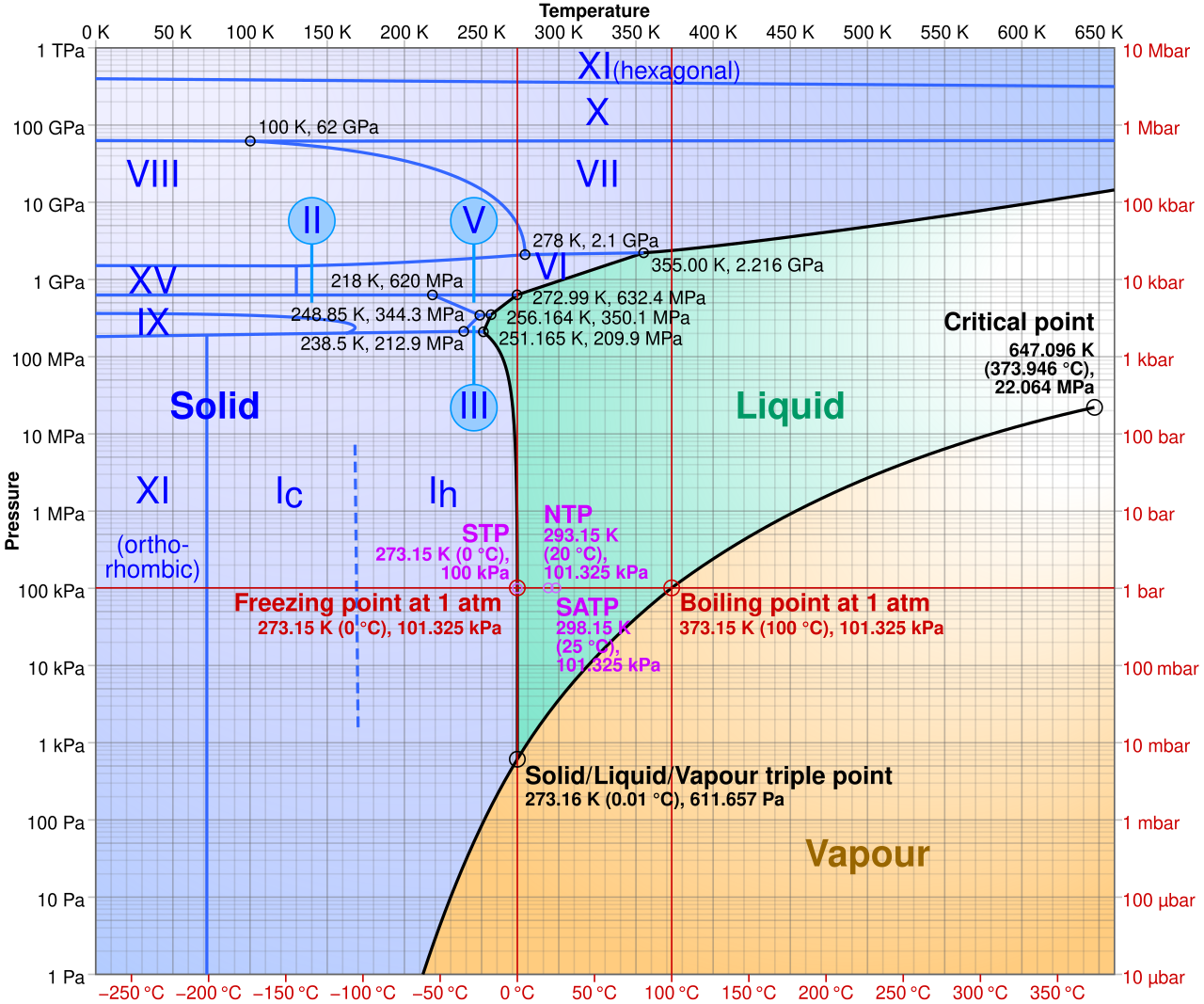}
    \caption{Phase diagram of water as a function of temperature and pressure \cite{water}.}
    \label{fig:water}
\end{wrapfigure}

\medskip\noindent
\textbf{Water} is one of the most abundant molecules in the universe and a prerequisite for life as we know it.
Hexagonal ice, known as ice-Ih, is the most common form and is produced by bulk nucleation of liquid water at ambient pressure.
There are surprisingly many other phases of ice in the temperature-versus-pressure phase diagram of water (Fig. \ref{fig:water}) \cite{hermann2012,rescigno2025}. Studying the lattice instabilities \cite{strassle2004} and phonon softening leading to these structural phase transitions is crucial in order to gain a better understanding of the properties of ice at extreme pressures in the solar system, like on Uranus, Neptune, Europa, and Ganymede. However, such studies are limited by the small volumes of material obtainable in pressure cells, and is often limited to vibrational dynamics \cite{li1996,koza2008}.

\begin{wrapfigure}{l}{0.3\textwidth}
    \includegraphics[width=0.3\textwidth]{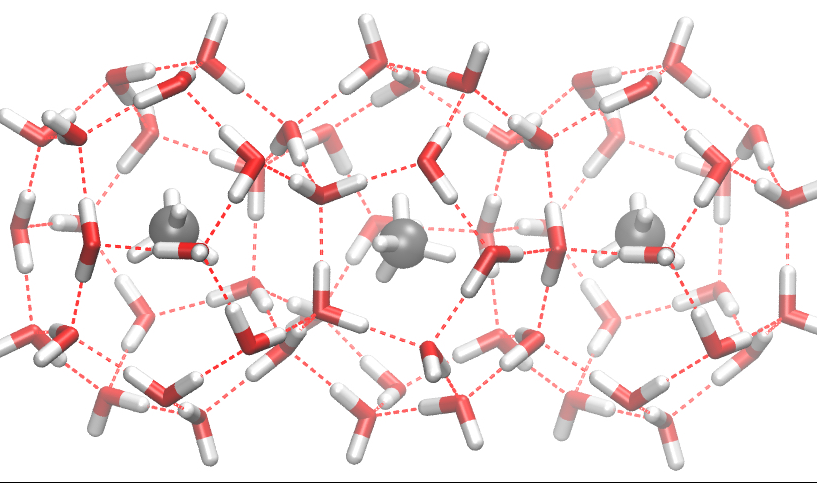}
    \caption{Methane trapped in a water cage \cite{nist}.}
    \label{fig:clath}
\end{wrapfigure}

Clathrates are crystalline inclusion compounds, where a guest molecule can be trapped in a chemical cage, often comprised of water (Fig. \ref{fig:clath}) \cite{sloan2003,medeiros2020}. The trapped molecule can be, e.g., methane, carbon dioxide or hydrogen, but it is not ionically bonded to the host lattice, which is stabilised through van-der-Waals interactions.
In many cases, such clathrates are formed under pressures of $0.1-3$ GPa and are fascinating objects of study as they are technically and geologically relevant as candidate gas storage materials and because they form naturally in comets and deep ocean sediments, respectively. The guest molecule is weakly bound to its surroundings and gives rise to interesting low-energy dynamics such as rattler modes, rotational tunnelling and hindered rotations \cite{goto2004,schaack2018} which are highly non-trivial to model using conventional DFT methods.
The study of these modes are often ${\bf Q}$-integrated \cite{ulivi2007,farrando-perez2022}, for the simple reason of limited sample mass.

Njord offers a unique opportunity to go beyond the mere density of states and gain detailed information about the phonon Q-dependence in systems like water and clathrates.

\medskip\noindent
\textbf{Barocalorics} refer to materials that exhibit the barocaloric effect, a thermodynamic phenomenon in which a material undergoes a reversible structural change when hydrostatic pressure is applied or removed. When pressure is applied, the material releases heat as its internal structure becomes more ordered; when the pressure is removed, the material absorbs heat and cools \cite{ismail2021}. Barocaloric materials are of growing interest because they enable solid-state refrigeration. Traditional refrigeration relies on gaseous refrigerants that can have significant environmental impacts; whereas barocaloric cooling uses solid materials potentially reducing greenhouse gas emissions and improving energy efficiency \cite{sun2025}. Recent research has identified a variety of promising barocaloric materials, including plastic crystals, polymers, and intermetallic compounds, some of which exhibit large entropy changes and temperature variations near room temperature \cite{poreba2023}.
With Njord it will be possible to study the pressure-induced changes of the vibrational spectra of these materials, which yields the smoking-gun evidence on the origin of the barocaloric effect.
Operando studies of the structure and dynamics of these materials during cooling cycles will be crucial to understand their driving mechanisms and hence to design barocaloric refrigerant devices for application.
Such experiments will be possible on Njord even on bulk barocaloric devices thanks to the well defined beam and spatial resolution on the order of mm.

\subsection{Quantum magnetism}

Quantum magnets are ideal testing grounds for many-body quantum models. The fingerprints of predicted ground states such as quantum-spin liquids and resonant-valence-bond states are found in the associated excitations \cite{savary2016,knolle2019} and here inelastic neutron scattering is one of the most powerful tools to identify these signatures. Moreover, measures such as Quantum Fischer Information can be used to directly extract entanglement measures from inelastic neutron scattering measurements and thereby quantify the ``quantumness'' of a given system \cite{laurell2025}.

In recent years the study of quantum effects and low-dimensionality in combination with magnetic frustration has intensified. This is driven by the sophistication and increased user-friendliness of theoretical tools such as the density-matrix-renormalisation-group and infinite-projected-entangled-pair-states methods in combination with developments in quantum information and quantum computing technologies. However, on the experimental side, and in particular with respect to neutron spectroscopy, a parallel advancement has been hindered by the inherently low neutron flux and hence need for large single crystals. To mention specific examples of systems with recent interest we highlight MnCr$_2$S$_4$, the most promising candidate material to realise a spin supersolid, and Na$_3$Co$_2$SbO$_6$, which is predicted to host a quantum spin liquid under pressure.

\begin{wrapfigure}{l}{0.7\textwidth}
    \includegraphics[width=0.7\textwidth]{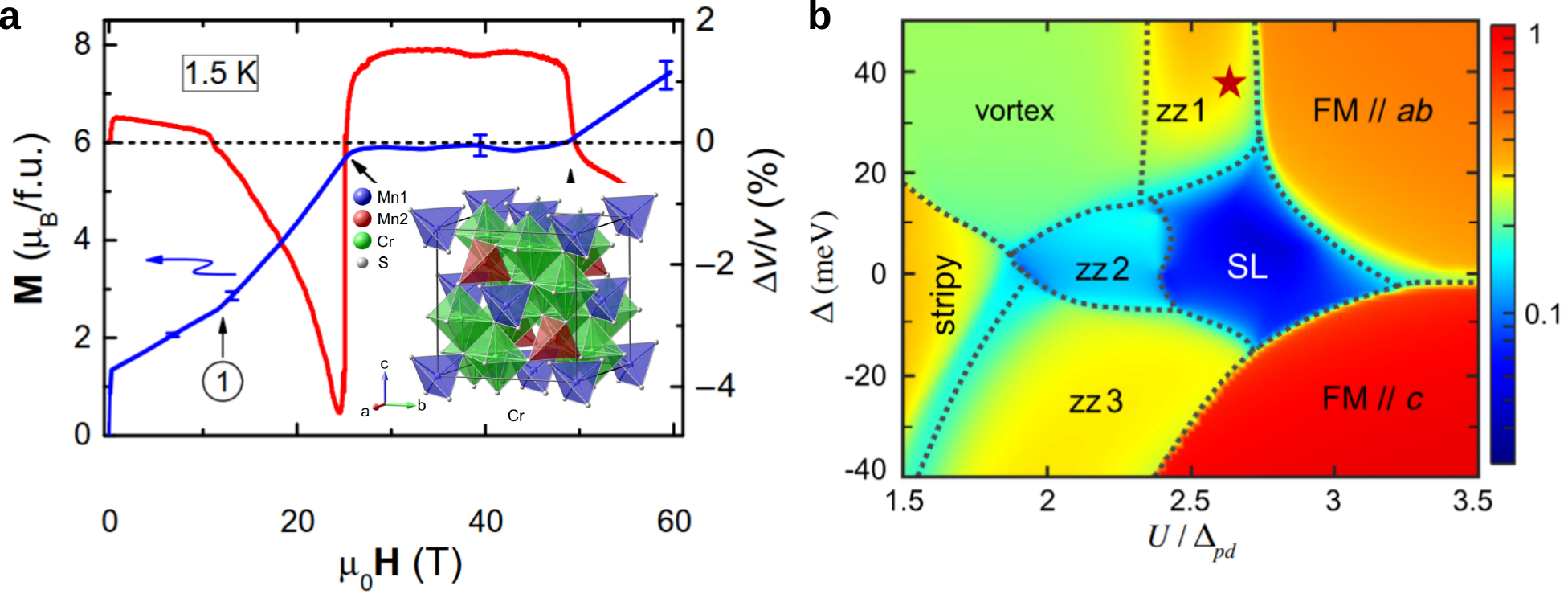}
    \caption{\textbf{a} Magnetisation curve of MnCr$_2$S$_4$ with its large plateau at $25-50$ T \cite{tsurkan2017}. Crystal structure is shown in the inset \cite{duc2024}.
    \\
    \textbf{b} Predicted phase diagram for the Kitaev honeycomb with the quantum spin liquid in dark blue and the ambient-pressure position of Na$_3$Co$_2$SbO$_6$ marked with the red star \cite{liu2020}.}
    \label{fig:quantum}
\end{wrapfigure}

\medskip\noindent
MnCr$_2$S$_4$ is a magnetic spinel compound that has attracted attention because it may host what is known as a spin supersolid phase \cite{plumier1980,tsurkan2017,ruff2019,duc2024}. This is an exotic quantum magnetic state that simultaneously exhibits rigid, ordered magnetic structures (solid) and superfluid-like dissipationless transport of magnetisation (superfluid) \cite{matsuda1970,fischer1973,batista2007,ueda2010}. It is characterised by coexisting longitudinal and transverse spin orders which appear to be present in MnCr$_2$S$_4$ for magnetic fields above and below a pronounced magnetisation plateau spanning the range $\sim25-50$ T, Fig. \ref{fig:quantum}a.
The magnetic structure in the lower-lying of the proposed spin supersolid phases ($11-25$ T) has been studied with neutron diffraction.
MnCr$_2$S$_4$ crystals grow on the order of $1-2$ mm in size making inelastic studies presently impossible; with the spin Hamiltonian thus undetermined the spin supersolid state is therefore unconfirmed.

\begin{sloppypar}
\medskip\noindent
    Na$_3$Co$_2$SbO$_6$ is a layered honeycomb cobaltate and a candidate platform for frustrated magnetism with Kitaev\--type interactions. Unlike conventional Heisenberg interactions that are isotropic, Kitaev interactions depend on the bond direction. The primary interest in these interactions is that they can stabilize a Kitaev quantum spin liquid, a topological state of matter with fractionalized excitations. At ambient conditions, Na$_3$Co$_2$SbO$_6$ orders antiferromagnetically but experiences significant frustration \cite{li2022,gu2024,hu2024} and theoretical and experimental work suggest that application of pressure will bring the material into the quantum-spin-liquid regime, Fig. \ref{fig:quantum}b \cite{liu2020,poldi2025}. While it is possible to grow relatively large Na$_3$Co$_2$SbO$_6$ crystals, single-domain or de-twinned samples rarely exceed $2-3$ mm sizes. Again, inelastic neutron scattering experiments are currently impossible for this size of samples but would be the pivotal probe for confirming the Kitaev quantum spin liquid in a real material.
\end{sloppypar}

\medskip\noindent
Njord addresses all of the above challenges (small samples, pressure and magnetic field). The extreme focusing of the neutron beam ensures a concentration of flux in a small area and naturally comes with a moderate $Q$ resolution. Typical excitations in quantum magnets, however, have little or no dispersion or form continua, meaning that the $Q$ resolution of Njord is more than sufficient to characterise these types of excitations.

\subsection{Key scientific drivers}
The combination of Njord and Remora encompasses almost the entirety of scientific drivers in cold neutron spectroscopy today.
Spanning from soft matter, through energy materials, to functional materials as well as fundamental planetary and quantum science; the experimental capability space covered by Njord and Remora is vast.
Below, we focus on three overarching categories:

\medskip\noindent
\textbf{Spintronics and quantum computing} hold perspectives to revolutionise information technology. The field of spintronics aims to harness spin-polarised currents and topological magnons to create logic devices and storage with almost no heat dissipation. In practice, device cooling is the main factor limiting the density of computing power, and dissipationless devices would have immense technological and environmental impact.
Quantum computing, on the other hand, aims to solve certain problems that are intractable for classical computers by harnessing the quantum phenomena of superposition and entanglement. If realised, it could accelerate the discovery of new materials and drugs, optimise complex systems like energy grids and logistics, and improve climate and financial modelling.
A fundamental understanding of altermagnetism, topological magnetism and quantum entanglement in magnetic materials are crucial to move these fields forward.

\medskip\noindent
\textbf{Energy sciences} cover everything from energy harvesting and storage to minimising energy consumption in our everyday lives when concerning transport and housing but also anything from computation, sensing or even entertainment.
Within this broad field, superconductors and MOF materials are particularly relevant.
Superconductors allow loss-less energy transport, high-efficiency public transport, energy storage and future magnets for fusion technology. On the other hand, breathing MOFs offer perspectives for selective carbon capture and photocatalysis, while clathrates enable safe storage of hydrogen and methane.
In such systems, the crystal lattice is not just a given, is has a concrete almost mechanical function, and making these materials functional requires an understanding of the collective lattice dynamics.
Njord will also allow to study diffusion processes by QENS with relaxation times $< 100$~ps.
This will allow, e.g., spatially resolved studies on the loading and unloading of hydrogen storage materials with mm spatial resolution.

\medskip\noindent
\textbf{Fundamental science} like phases of ice under high pressure, collective dynamics in MOFs and quantum magnetic phases remain central elements of the science conducted at Njord.

\subsection{Potential new science and new users}

Due to space constraints, this proposal has detailed only some of the new science enabled by Njord.
In particular the research fields of MOFs and organic superconductors will be newly accessible to neutron spectroscopy.
By expanding the realm of feasible experiments to these and other new science areas, new user bases across Europe will be brought into the fold of inelastic neutron scattering on collective excitations.

\section{Technical overview}
\begin{figure}[!h]
  \centering
  \includegraphics[width=0.99\linewidth]{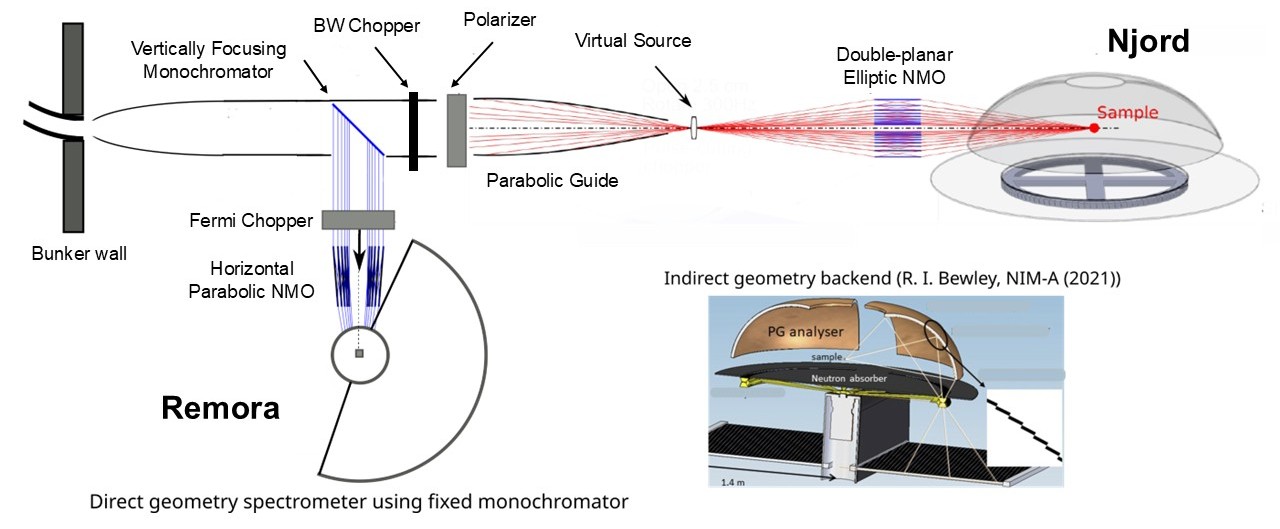}
  \includegraphics[width=0.99\linewidth]{ToF_Diagram.png}
  \caption{(Top) Overview of the instrument setup (not to scale). Upstream of the bandwidth chopper (BW) of Njord, a vertically-focusing monochromator feeds Remora. A parabolic trumpet with a slit creates a well-defined virtual source, mirrored by an NMO onto the sample position of Njord. (Bottom) ToF Diagram, where the $1.7$ Å bandwidth reaching Njord is shaded yellow to purple, while the green rays represent the neutrons that fulfil the $(002)$, $(004)$ and $(006)$ monochromator mosaic Bragg scattering for Remora.}
  \label{fig:CrudeOverview}
\end{figure}

Njord is an indirect-geometry ToF spectrometer designed for extreme flux on small samples. Remora is a symbiotic direct-geometry ToF spectrometer extracting neutrons from the same beamport and exploiting the excess wavelength band not used by Njord.
The overall concept is illustrated in \cref{fig:CrudeOverview}.

\medskip\noindent
\textbf{Njord} allows a continuous wavelength band on the sample, and employs a MUSHROOM-type crystal-analyser array, first put forward by R. Bewley from ISIS \cite{BEWLEY2021165077}. It accepts a very large solid angle of scattered neutrons. To achieve $2.5 \times 10^{10} \text{n/s/cm}^2$ on a \SI{3}{mm}$\times$\SI{3}{mm} beam spot, Njord makes use of Nested Mirror Optics (NMO) to focus the neutron beam onto the sample position \cite{HERB2022167154}. Such an optical component is composed of several short mirrors with coinciding focal points, which are stacked to manipulate the entire beam cross section along a short fraction of the total beam path. This design reduces geometrical aberrations and multiple reflections and, in contrast to conventional neutron focusing optics, does not require close proximity to the sample for a well defined neutron beam (more details about NMO are given in Sec. 3.2).

\medskip\noindent
\textbf{Remora} extracts distinct neutron wavelengths from the neutron guide upstream of the bandwidth chopper of Njord by means of a highly oriented pyrolitic graphite (HOPG) monochromator. A Fermi chopper controls the illumination time of the sample. The instrument features a compact secondary flight path of \SI{2.5}{m} to provide good energy resolution and maintain a large solid angle coverage. We match the higher-order reflections of the monochromator to the repeat wavelength of the chopper in order to increase the dynamic range by extracting two or more wavelengths from one ESS pulse. For Remora, an NMO will provide horizontal focusing while the monochromator focuses vertically.

\medskip\noindent

\subsection{Primary spectrometers}
A large phase-space volume is extracted from the bright ESS moderator using a neutron guide\footnote{We also consider using the NMO technology for optimising the initial neutron extraction following discussions with Peter Böni.}.
The divergence is transformed into a large cross section that can be transported efficiently.
The Remora monochromator extracts different Bragg reflections in the low-divergence region of the guide.
Further downstream, Njord accepts a bandwidth of \SI{1.7}{\angstrom} with $\lambda<$ \SI{4.7}{\angstrom}. The beam is then compressed once more into a bright virtual source which is imaged precisely by the NMO onto the sample position.
Our preliminary McStas simulations presented here were performed using the BIFROST and T-REX guide geometries for Njord and Remora respectively. These are not optimised for this design and therefore do not provide the ideal moderator extraction efficiency or divergence control expected from a dedicated guide tailored to this concept.

Remora is a multi-wavelength instrument utilising first- and higher-order reflections from a graphite monochromator, with a baseline configuration targeting \SI{4.8}{\angstrom} and \SI{2.4}{\angstrom} (\SI{3.55}{meV} and \SI{14.2}{meV}). Neutrons with $\lambda < 4.8$~\AA\ are for the most part transmitted through the graphite monochromator and utilised by Njord, taking advantage of the high transparency of graphite off the Bragg condition (exceeding 85\% transmission in this range). Even in a configuration where Remora employs the $(004)$ reflection at \SI{2.4}{\angstrom}, Njord can still fully exploit this wavelength band, as the second-order reflectivity remains below 30\% and any induced intensity variations are normalised. This shared use of phase space, combined with the possibility of integrating a pulse-shaping chopper (PSC) to enhance Njord’s resolution without significantly reducing flux on Remora, results in a well-balanced trade-off. This combination is particularly attractive for a next-generation ESS instrument which will complement and fortify the versatility of the first-generation instrument suite.

\subsection{Njord: Tiny samples in complex sample environments}

\begin{table}[b!]
\centering
\begin{minipage}[c]{0.55\linewidth}
 \vspace*{0pt}
 \captionsetup{width = 0.95\textwidth}
 \captionof{table}{Summary of specifications for Njord. Beam characteristics were simulated for a $2.45$--$4.25$~\AA\ bandwidth. Energy resolutions were calculated analytically and validated using McStas (\cref{fig:Njord_Secondary}).}
 \label{tab:Njord_Specs}
    \begin{tabular}{l l c}
        \multicolumn{3}{c}{Primary spectrometer}\\\hline
        \multicolumn{2}{l}{Beam spot size} & \SI{3}{mm}$\times$\SI{3}{mm}\\
        Flux - PSC parked & \multicolumn{2}{r}{$2.5 \times 10^{10} \text{ n/s/cm}^2$}\\
        \multicolumn{2}{l}{Divergence - horz. \& vert.} & $\pm 2\ \deg$\\\hline \vspace{2mm}\\
        \multicolumn{3}{c}{Secondary spectrometer}\\\hline
        \multicolumn{2}{l}{$E_f$ coverage} & \{3.4,\ 5\} meV\\
        \multicolumn{2}{l}{Out-of-plane coverage} & $\{15,\ -2.3\}\ \deg$\\
        & & $\{-0.4,\ 0.05\}$ \si{\angstrom^{-1}}\\
        \multicolumn{2}{l}{In-plane coverage}  & $\{\pm5,\ \pm140\}  \deg$ \\
        \multicolumn{2}{l}{Analyser solid angle coverage}    & 1.4 Sr\\
        \multicolumn{2}{l}{Energy transfer coverage} &  $\{-1.3, \ 30\}$ meV  \\
        \multirow{2}{8em}{Energy resolution (0 meV)} & PSC parked & \SI{190}{\micro eV}\\
                          &  PSC 0.5 ms & \SI{42}{\micro eV}\\
        \multirow{2}{8em}{Energy resolution (10 meV)} &  PSC parked & \SI{0.9}{meV}\\
                                   &  PSC 0.5 ms & \SI{0.12}{meV}\\
                                   \hline
\end{tabular}
\end{minipage}
\hspace{5mm}
\begin{minipage}[c]{0.37\linewidth}
    \vspace{0pt}
    \includegraphics[width=0.8\linewidth]{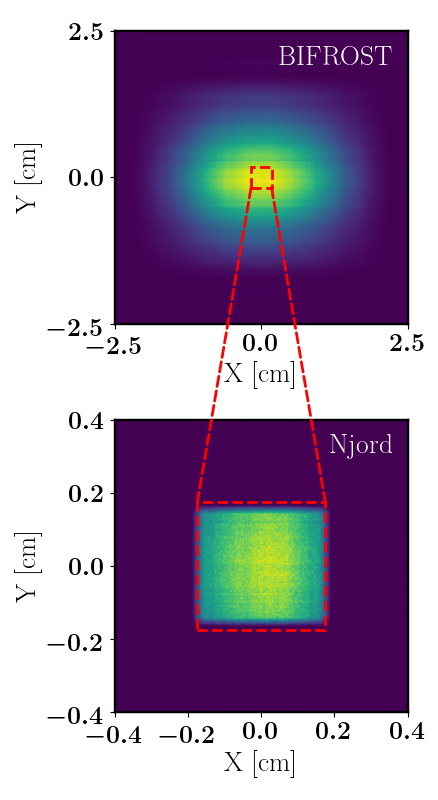}
    \captionsetup{width = 0.9\textwidth}
    \captionof{figure}{Comparison between simulated beam spots for BIFROST and Njord. The dashed red lines indicate the $3.5 \times 3.5$ \si{mm^2} beam area used for performance estimates.}
    \label{fig:beamspot}
\end{minipage}
\end{table}

Njord shifts the limits for neutron spectroscopy towards the sub-mm size samples. This extends the applicability of inelastic and quasi-elastic neutron scattering to research fields where only tiny samples may be provided. Furthermore, it enables the use of sample environments with severe restrictions on the sample volume such as high pressure devices.

The NMO provides ultimate flux in a sharply defined illumination area to reduce the experimental background without sacrificing intensity by the use of apertures or slits.
The concept relies on a set of compact, elliptically shaped, nested supermirrors that accept a large divergence from a virtual source and refocus it onto the desired area with high efficiency while preserving phase space. The reported experimental results, obtained with a small prototype elliptic NMO equipped with non-polarising supermirrors, have demonstrated a brilliance transfer for small samples of 72\%. The demonstration of sharp focusing of sub-mm wide beams was limited only by the spatial resolution of the detector of $2.5$ mm \cite{HERB2022167154}. An elliptic NMO built for a MIEZE focusing setup at MIRA at FRM II was tested at BOA at PSI with a detector having a resolution well below 0.1 mm \cite{Boeni2026}. This experiment further demonstrated the excellent focusing properties of the NMO. Moreover, NMO can efficiently provide a polarised beam when using polarising mirror coatings, meaning that upgrading Njord for polarisation analysis is straight forward in the future.

The guide focuses the beam onto a virtual source slit located several meters from the sample. A compact set of NMOs then images this directly onto the sample position with minimal smearing (Fig. \ref{fig:beamspot}). Our preliminary simulations, utilising the BIFROST extraction and transport guide, indicate that this setup can achieve a flux of approximately $2.5 \times 10^{10} \text{ n/s/cm}^2$ at 2 MW (Table \ref{tab:Njord_Specs}). This represents a 3- to 4-fold increase over BIFROST, with the same extraction system.

The aggressive focusing of the beam comes with a relatively large divergence of $\pm2^{\circ}$. Accepting this as a design premise allows the secondary spectrometer to utilise horizontal focusing while covering the entire scattering angle range in a single setting. Although this results in a scattering angle resolution of approximately $1.5^{\circ}$ or more, it remains largely compatible with our incoming divergence. With this resolution, the MUSHROOM design \cite{BEWLEY2021165077} serves as a nearly optimised template for analysing the scattering neutrons. It employs horizontal wide-angle focusing alongside vertical angular coverage. This configuration enables a single analyser bank to span an extensive angular range, thereby expanding the $Q$-range and facilitating the recording of full 4D $({\bf Q}, E)$ volumes. A beryllium filter with collimation between sample and analyser will suppress second order reflection off the HOPG, and a holistic optimisation of the analyser filter should be performed in the future.

\begin{figure}[t!]
  \centering
  \includegraphics[width=0.475\linewidth]{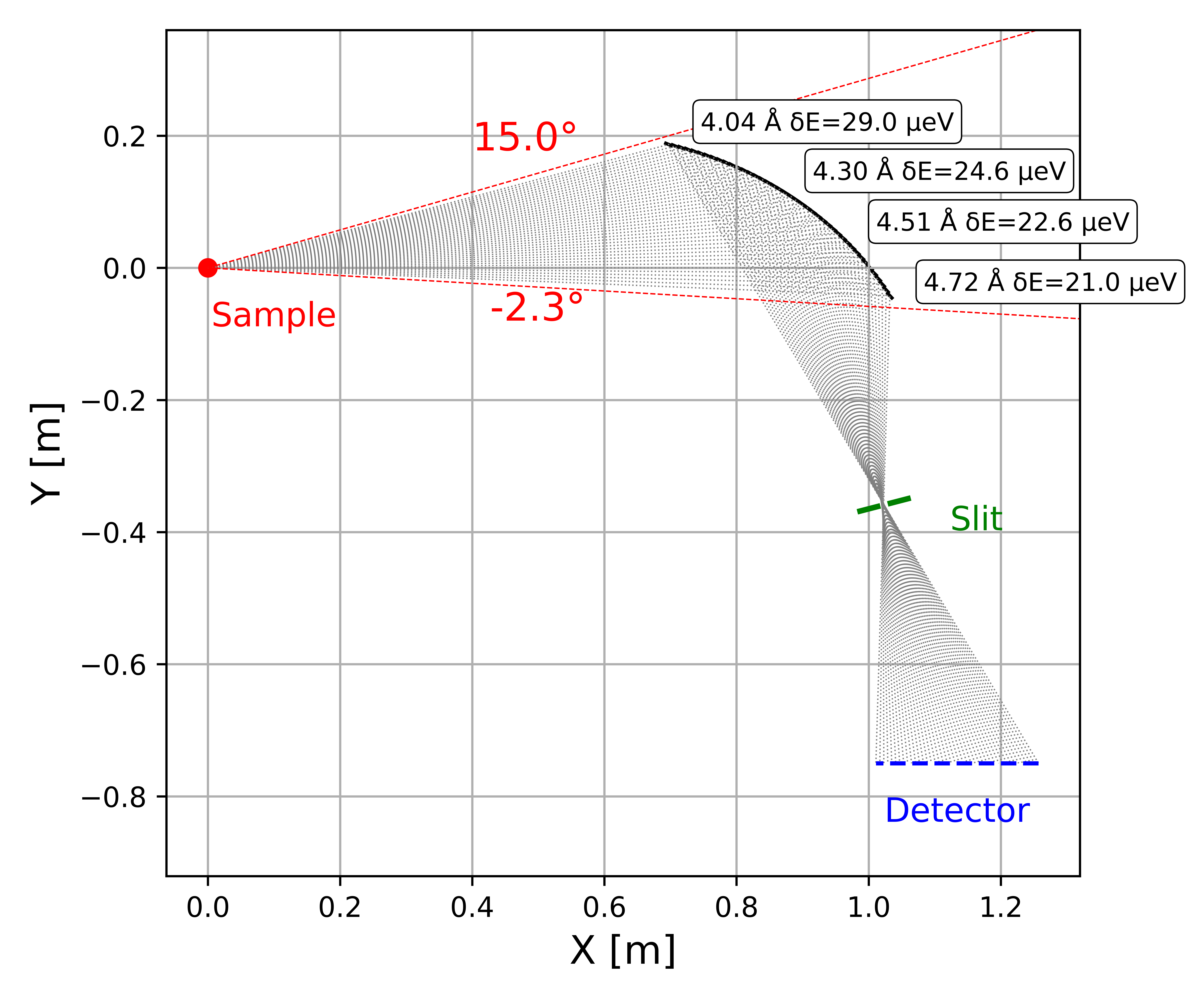}
  \includegraphics[width=0.475\linewidth]{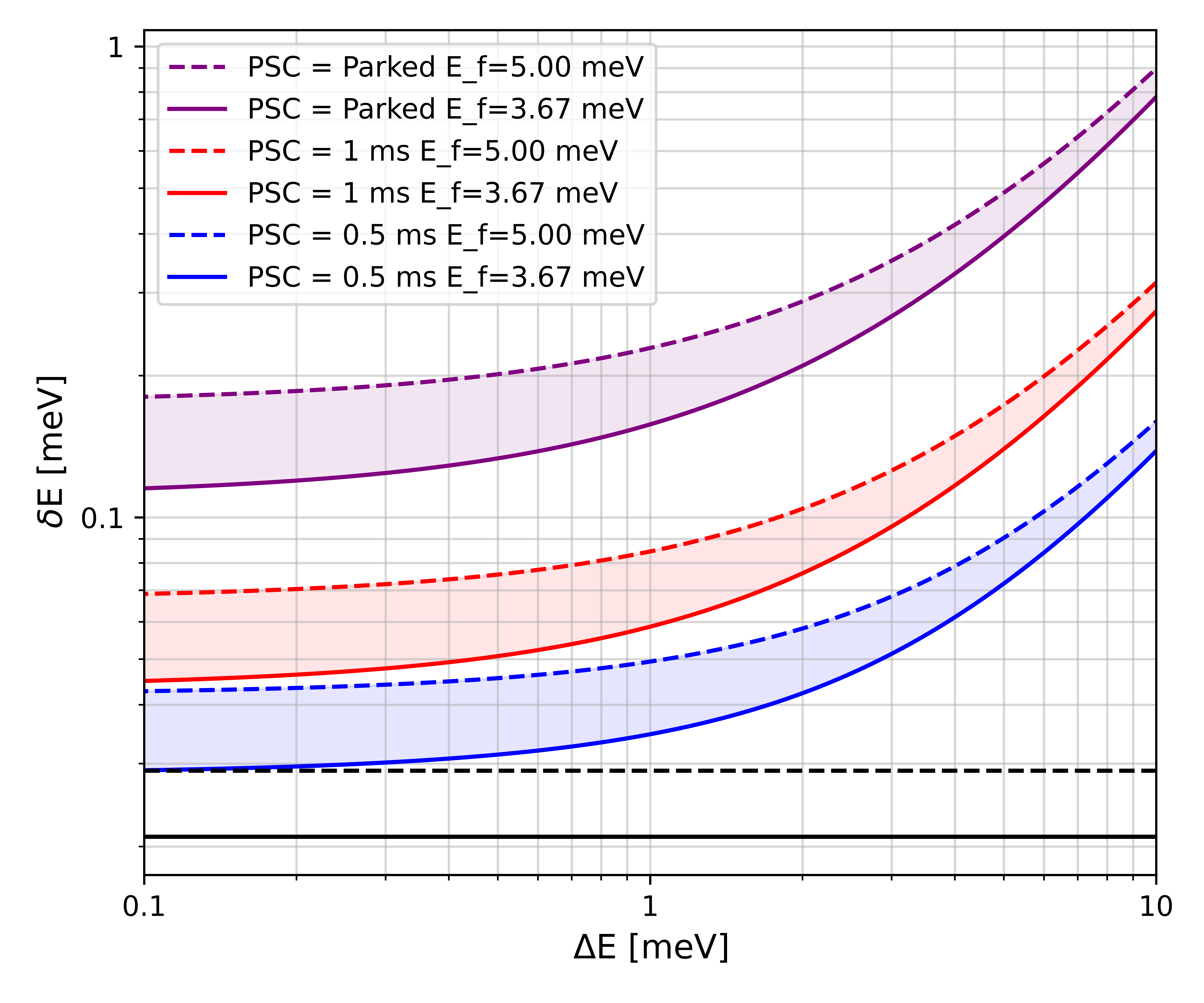}
  \caption{Illustration of the Njord secondary spectrometer (left) alongside energy-resolution curves for different PSC opening times (right). The secondary spectrometer configuration shows a single analyser array, spanning 17.3$^{\circ}$ out-of-plane coverage. The resolution curves show the resolution for 3 PSC opening times (fully parked PSC, 1 ms opening time, and 0.5 ms opening time) for the lowest and highest final energy analysers. The horizontal black lines show the resolution of the secondary spectrometer for the lowest (full) and highest (dashed) final energies.}
  \label{fig:Njord_Secondary}
\end{figure}

A preliminary instrument design has been simulated and is illustrated in \cref{fig:Njord_Secondary} together with the corresponding final energies, and the resulting energy resolution as a function of vertical scattering angle. We have adopted an asymmetric design that probes vertical scattering angles from +15$^{\circ}$ to -2.3$^{\circ}$. This configuration enables the use of a magnet optimised for vertical angular coverage, commonly employed on, e.g., IN5. The energy resolution can also be seen in \cref{fig:Njord_Secondary}. Without pulse shaping it matches typical resolution settings of triple-axis spectrometers using double focusing. Even with modest pulse shaping the energy resolution is comparable to existing direct-geometry ToF spectrometers. In combination with the large solid angle coverage Njord is also very well suited for QENS studies on $< 100$-ps timescales.

Njord presents a significantly more open scattering geometry compared to most other indirect-geometry ToF spectrometers due to the MUSHROOM analyser. Despite the clear advantages of the increased out-of-plane coverage it also comes with its own challenges regarding tracking scattering neutrons. This is partially mitigated by the use of focusing analyser elements that direct the scattered neutrons onto a circular slit between analyser and detector, thereby restricting the effective field of view of each detector pixel.
In this configuration, the detector bank comprises tubes arranged radially with respect to the sample rotation axis. Different final energies are sampled as a function of position along each tube, exploiting the prismatic concept \cite{cameapaper,Birk2014}. This not only improves the energy resolution, but also allows for an increased acceptance of the analysers and provides high intensity at the detector.

\begin{figure}
    \centering
    \includegraphics[width=\linewidth]{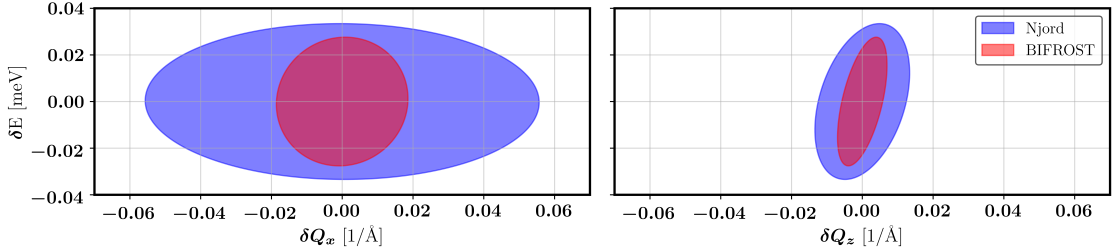}
    \caption{Comparison of FWHM resolution ellipses between Njord (blue) and BIFROST (red) for 1 ms PSC opening time, at 0 meV energy transfer, at a $45^\circ$ in-plane scattering angle, and for the 5 meV analyser in both cases. Left panel shows the resolution perpendicular to the incident beam in the plane, and right panel shows the resolution parallel to the incident beam in the plane.}
    \label{fig:Njord-Q-Res}
\end{figure}

The $Q$-resolution of the instrument is illustrated in \cref{fig:Njord-Q-Res}. It can be seen how the resolution is affected by the large incident divergence. To allow for more flexibility, as some experiments may require a better $Q$-resolution, a slit could be placed in front of the NMO. This would allow the instrument to dynamically change the incident divergence, separately for the vertical and horizontal directions.
Such a reduction in divergence would also reduce incident flux without impact on illuminated spot size.

\subsection{Remora: A hybrid spectrometer}

\begin{table}[b!]
    \centering
    \begin{minipage}[c]{0.55\linewidth}
    \vspace{0pt}
    \captionsetup{width = 0.95\textwidth}
    \captionof{table}{Summary of specifications for Remora as determined by preliminary McStas simulations using a T-REX guide.}
    \label{tab:REMORA_Specs}
    \begin{tabular}{l l c}
        \multicolumn{3}{c}{Primary spectrometer}\\
        \hline
        \multicolumn{2}{l}{Beam spot size} & \SI{2.15}{cm}$\times$\SI{1.87}{cm}\\
        Divergence & horizontal & $2.0\ \deg$\\
                   & vertical   & $2.6\ \deg$\\
        \multicolumn{2}{l}{$\delta \lambda / \lambda$ for 4.8 Å} & 2.3\%\\ \hline \vspace{2mm}\\
        \multicolumn{3}{c}{Secondary spectrometer}\\
        \multicolumn{3}{c}{(Fermi chopper frequency \SI{112}{Hz})}\\\hline
        & $\lambda=$\SI{2.4}{\angstrom} & $\lambda=$\SI{4.8}{\angstrom}\\
        Flux $\text{ n/s/cm}^2$ & $9 \times 10^{4}$  & $3 \times 10^{5}$ \\
        \vspace{0.1mm}\\
         & @$\Delta E=$\SI{7.1}{meV}   & @$\Delta E=$\SI{0}{meV}\\
         Energy Resolution & \SI{380}{\micro eV} &  \SI{150}{\micro eV}\\
        \hline
    \end{tabular}\vspace{1mm}\\
\end{minipage}
\hfill
\begin{minipage}[c]{0.4\linewidth}
    \vspace{0pt}
    \includegraphics[width=\linewidth]{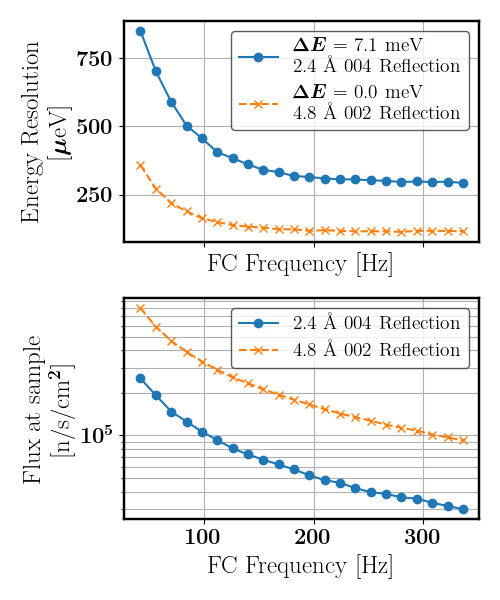}
    \captionsetup{width = 0.95\textwidth}
    \captionof{figure}{(Top) Elastic energy resolution for 4.8 Å, and 7.1 meV energy resolution for 2.4 Å scattering processes on REMORA, for different Fermi chopper frequencies. (Bottom) Flux at the sample position for different Fermi chopper frequencies.}
    \label{fig:remoraresolution}
\end{minipage}
\end{table}

As Njord maximally uses a certain band of neutrons (\SI{1.7}{\angstrom}), the excess spectrum resulting from using the full ESS pulse can be used to feed a symbiotic second instrument situated on the same guide. We propose to introduce a HOPG monochromator \SI{70}{m} from the moderator, which is mostly transparent for all wavelengths that are not Bragg diffracted.
This monochromator can be installed upstream of the bandwidth chopper system for Njord. Setting a maximum incident wavelength on Njord to \SI{4.7}{\angstrom}, Remora's fundamental wavelength is proposed to be \SI{4.8}{\angstrom}.

The use of multiple incident energies on the sample position (also known as repetition rate multiplication \cite{Mezei:2002uq}) is readily increasing the efficiency of direct-geometry instruments at low-frequency pulsed sources. Remora extends this concept to a hybrid spectrometer using the $(002)$, $(004)$, and $(006)$ Bragg reflections of the HOPG monochromator. Thanks to the pulsed nature of the source, the different HOPG harmonics are well separated in time. This allows the simultaneous recording of several spectra collectively covering a large dynamic range ($0.1 \mathrm{meV} < \hbar \omega < 50 \mathrm{meV}$) to study excitations on multiple time scales.

Placing the monochromator at a distance $L_\mathrm{mono} = 70$~m from the source, we gain sufficient room to the neighbouring instruments and eliminate overlap from adjacent ESS pulses.
As the common beam transport with Njord prevents focusing the beam onto a virtual source, we envision vertically focusing using the monochromator, while horizontal focusing will be achieved using a compact NMO.

The secondary energy resolution of this instrument will be mainly controlled and adjusted by a Fermi chopper, as seen in \cref{fig:remoraresolution}, which defines the illumination time of the sample. This chopper will be positioned between the monochromator and sample such that it minimises the time spread resulting from the finite wavelength resolution of the monochromator.
The secondary spectrometer is kept rather compact to allow a large solid-angle coverage at an acceptable price for the detector. Our preliminary optimisation yielded here a sample-to-detector distance of $2.5$ m. This is similar to the ILL hybrid spectrometers, Panther and Sharper.
As seen in \cref{tab:REMORA_Specs}, the elastic \SI{4.8}{\angstrom} resolution becomes constant for chopper frequencies $> 150$ Hz. At this frequency even the unoptimised neutron transport yields a monochromatic flux of $2\times 10^5$ n/s/cm$^2$, i.e. comparable to world-class ToF spectrometers today.

\section{Use of the ESS long-pulse source}
The Njord concept is designed to make full use of the ESS pulse, by virtue of being a flux-optimised instrument operating close to the feasibility limit. The PSC enables a limited flux–resolution trade off, with a minimum pulse width of 0.7 ms. While the flexibility of the long ESS pulse is highly advantageous, aggressive pulse shaping effectively converts ESS into a low-frequency short-pulsed source in terms of time-integrated brilliance. ESS therefore delivers its greatest strengths when employed in indirect geometry, where the full source brilliance can be harvested. Njord applies this principle to its extreme.

The long neutron pulse provides the large phase space to serve two instruments at one beamline of a spallation source and hence makes Remora possible.
Together, the two instruments maximise utilisation of the ESS brightness.

\section{Sample environment and laboratory access}

In the field of experimental quantum magnetism and traditional hard condensed matter physics, the existence of relevant model materials relies on inorganic chemical synthesis. In many cases, like for the two examples in Sec. 2.4, only mm-sized single crystals can be grown. Moreover, when larger samples are possible, they are often twinned, meaning that the interpretation of neutron spectra becomes exceedingly difficult. Even in the few cases where large single-domain crystals exist, the interaction parameter regime may still be somewhat away from what is desired. Therefore, to manipulate the magnetic ground state as well as to study critical phenomena, the application of external stimuli is crucial. Here hydrostatic (isotropic) pressure is one of the most important tools. The same may be said for uniaxial pressure and in addition this method may be used to de-twin samples. In both cases, $P = F/A$, where mechanical limitations in applied force mean that higher pressures are obtained only by decreasing sample size, which, in turn, is incompatible with today's neutron instrumentation.

Apart from pressure, magnetic field is an important external control parameter for studying phase transitions but also for determining spin Hamiltonians in the fully polarised (magnetised) state. Here both vertical and horizontal fields are needed depending on the spin orientation and propagation vector in question.

Njord is designed to excel as an extreme-environment spectrometer and will therefore benefit from a broad and flexible sample-environment portfolio. As Njord would be commissioned at a time when ESS is a fully mature facility, high-field magnets, dilution refrigerators, and standard pressure cells will be available. The development of instrument-specific pressure environments may minimise background. Especially sintered diamond-anvil pressure cells become an exciting avenue, since they offer a larger pressure range, but these have so far only been used for neutron diffraction (e.g. at SNAP \cite{Haberl2023}) and not yet for neutron spectroscopy.

Njord is well matched to horizontal-field magnets, inherently limiting horizontal scattering-angle coverage while permitting a relatively large vertical opening. Njord exploits this geometry efficiently, providing gapless coverage of the accessible scattering angles together with sensitivity to out-of-plane scattering.

\section{Proposed location of the instrument at the ESS facility}

In order of exploit the full ESS pulse and retain a relaxed resolution, Njord needs to be around $160$ m long, the natural length of a PSC instrument at the ESS. This means positioning Njord in a new instrument hall, not currently in place but possible options are currently being discussed. We mention that Njord could be a potential instrument to benefit from the high-brightness cold moderator proposed by L. Zanini and co-proposers.

For Remora, at 70 meters from the moderator, there are roughly 7.5 m from the centre of the guide to the start of the shielding blocks at the adjacent instrument in the most favourable positions. With a detector 2.5 m away from the sample position, Remora requires at least 7 m to the next beamline in order to allow enough space for shielding.

\section{Scientific gap analysis and comparison to other instruments}

\subsection{Njord: More flux, smaller samples, extreme conditions}
Njord offers, for the first time, the opportunity to study condensed matter dynamics on sub-mm-sized samples. Being a relaxed-resolution, wide $Q$-range cold spectrometer, Njord would take its place in a well-established scientific ecosystem. It is not a new instrument concept looking for a science case but a well established method being pushed to the extreme and thereby opening up neutron spectroscopy for new research fields. The science case has already been described Sec. 2, and the key point to mention here is the instrument uniqueness: Njord will be the only place in Europe, perhaps even in the world, where the dynamics of some of these materials can be studied. Njord does not simply address an existing gap; it redefines the operational space in which neutron spectroscopy serves, allowing new problems to be discovered and studied. As compared to BIFROST, Njord is not a competitor but an enabler, designed to perform low-resolution experiments that will be impossible even there.

One type of experiment especially promising on Njord is powder spectroscopy on small samples. This is currently not feasible on direct geometry spectrometers, and BIFROST is suboptimal as it is designed for high resolution in a single scattering plane. The large spatial angle covered by Njord is optimal for such studies.

\subsection{Remora: More is more}

The use of a single beam port by more than one instrument has up to now not been realised at pulsed neutron sources. Only the long ESS pulse provides enough time to extract distinct spectral ranges that serve the need of the individual instrument without severely hampering its compagnon. On the capability-capacity spectrum, Remora addresses the latter. Remora will enable experiments like those feasible today on LET, IN5, CNCS, CAMEA, DCS, AMATERAS, and similar instruments. Within the ESS instrument suite, it tackles similar questions as CSPEC, T-REX and BIFROST. The performance of Remora will lie somewhere in between LET and IN5, instruments that have had oversubscription factors over 4 for decades. The demand for cold neutrons at all facilities is very high with the oversubscription on LET reaching a factor of 7 after its shutdown for chopper and guide maintenance in 2023/2024. Having a third cold spectrometer at ESS with a high signal-to-noise ratio, competitive flux and the ability to perform experiments which do not require the full capabilities of CSPEC or T-REX will increase the throughput for the ESS as a facility.

Another important aspect of Remora is for training the next generation of neutron users and instrument scientists attending schools such as HERCULES. For this purpose, it is important to ensure an instrument similar to the old IN3 triple-axis spectrometer at ILL is available and Remora fills this role.

\section{Conclusion}
The main motivation for building ESS was to shatter the boundaries of the feasible and become the main facility for neutron scattering in Europe.
Our proposal for Njord and Remora directly targets both of these ambitions.
Njord will increase collective excitation count rates by almost an order of magnitude, opening up new science cases previously inaccessible in the first ESS suite.
Remora, on the other hand, will provide capacity using a conventional design and allowing the excellent science being done today to remain available beyond 2035.

\addcontentsline{toc}{section}{References}
\printbibliography[title={References}]

\end{document}